 \documentclass{emulateapj}

\shorttitle{{\it SPEAR} Instrumentation}
\shortauthors{EDELSTEIN, et al.}

\begin{document}

\title{The SPEAR Instrument and On-Orbit Performance}

\author{Jerry Edelstein\altaffilmark{1}, Eric Korpela\altaffilmark{1}, Joe Adolfo\altaffilmark{1}, Mark Bowen\altaffilmark{1}, Michael Feuerstein\altaffilmark{1}, Jeffrey Hull\altaffilmark{1}, \\ Sharon Jelinsky\altaffilmark{1}, Kaori Nishikida\altaffilmark{1}, Ken McKee\altaffilmark{1}, Peter Berg\altaffilmark{1},  Ray Chung\altaffilmark{1}, Jorg Fischer\altaffilmark{1}, \\
Kyoung-Wook Min\altaffilmark{2}, Seung-Han Oh\altaffilmark{2}, Jin-Guen Rhee\altaffilmark{2}, Kwangsun Ryu\altaffilmark{2}, Jong-Ho Shinn\altaffilmark{2}, \\
Wonyong Han\altaffilmark{3}, Ho Jin\altaffilmark{3}, Dae-Hee Lee\altaffilmark{3}, Uk-Won Nam\altaffilmark{3}, Jang-Hyun Park\altaffilmark{3}, Kwang-Il Seon\altaffilmark{3}, In-Soo Yuk\altaffilmark{3}}

\affil{$^1$Space Sciences Laboratory, University of California, Berkeley, CA 94720}
\affil{$^2$Korea Advanced Institute of Science and Technology, Dajeon, Korea 305-701}
\affil{$^3$Korea Astronomy Observatory, Dajeon , Korea 305-348}

\begin{abstract}

The $\it SPEAR$ (or $``FIMS''$) instrumentation
has been used to conduct the 
first large-scale spectral mapping of diffuse cosmic far ultraviolet 
(FUV, 900-1750 {\AA}) emission, including
important diagnostics of interstellar hot (10$^{4}$ K -- 10$^{6}$ K) 
and photoionized plasmas, H$_{2}$, and dust scattered starlight.
The instrumentation's performance has
allowed for the unprecedented detection of astrophysical
diffuse far UV emission lines.
A spectral resolution of $\lambda/ \delta \lambda \sim$ 550
and an imaging resolution of 5'
is achieved on-orbit in the 
Short (900 -- 1175 {\AA})
and Long (1335 -- 1750 {\AA})
bandpass channels within their respective
7.4$^{\circ}$ x 4.3'
and
4.0$^{\circ}$ x 4.6'
fields of view.
We describe the {\em SPEAR} imaging spectrographs,
their performance,
and the nature and handling of their data.

\end{abstract}

\keywords{ultraviolet: ISM, instrumentation: spectrographs
}

\section{Introduction}

The Spectroscopy of Plasma Evolution from Astrophysical Radiation 
(hereafter {\em SPEAR} and $a.k.a ``FIMS''$) 
payload is surveying the sky for cosmic FUV emission. 
The {\em SPEAR} mission, launched 27 Sep. 2003,
its science objectives, and the mission profile and performance
are described in \citet{je:05apj_mission}.
{\em SPEAR's} bandpass (\ensuremath{\lambda}\ensuremath{\lambda} 900 -- 1750 {\AA}), 
spectral resolution ($\lambda/ \delta \lambda \sim$550), 
large field of view (7.4$^{o}$ x 4.5'), and imaging resolution 
(\ensuremath{\sim}5') facilitate the measurement 
of energetic interstellar gas while 
rejecting air-glow emission and stellar contamination that have 
plagued earlier measurements attempts. 
This paper describes the {\em SPEAR} instrument, 
its on-orbit performance, and the basic processing of instrument data.
Measurements of faint diffuse FUV emission lines are a
difficult undertaking wherein instrumental and systematic errors can dominate 
the results. Therefore, we discuss instrumental facets in  detail
in order to allow users to understand the nature and quality of {\em SPEAR} data.
Further technical 
description regarding various aspects of the {\em SPEAR} instrument
are found in the literature
\citep{ je03.spie.miss, korp03.spie.inst, nam03.spie.detelec, nam02.jass.det, ryu03.spie.opt}.

\section{The Spectrographic Method}

The {\em SPEAR} instrument consists of dual imaging spectrographs optimized 
for the detection of faint diffuse FUV radiation. The two spectral 
channels are designated as the ``Short'' (\ensuremath{\lambda}\ensuremath{\lambda} 
900 -- 1175 {\AA}) and ``Long'' (\ensuremath{\lambda}\ensuremath{\lambda} 1335 -- 1750 {\AA}) 
bandpasses. The bands were chosen to include
astrophysically important emission lines from abundant ionic 
species while avoiding instrumental contamination by the intense 
H {\sc I} $\lambda$ 1216 {\AA} and 
O  {\sc I} $\lambda$ 1304 {\AA} geocoronal radiation. 
Each channel uses an optical configuration
(see Fig. 1)
consisting of a collecting mirror, a slit, a diffraction grating, and a detector.
The unique two-element f/2.2 optical system has 
an off-axis parabolic-cylinder collecting mirror that
focuses plane-parallel light to a slit.
Cylindrical radiation from the slit then strikes a 
constant-ruled grating with an elliptical figure
that corrects on-axis aberrations to the third order. 
Diffracted light is imaged as a spectrum 
on a planar position sensitive photon counting detector. 
Radiation from off-plane angles is imaged along the detector 
perpendicular to the dispersion plane, analogous to a slit-imaging 
spectrograph. The 
cylindrical-source scheme doubles the usable 
imaging angle in comparison with classical point-source spectrographs. 
The result is a large solid angle -- collecting area product, 
a determining factor for diffuse source sensitivity. The design 
is an extension of the \textit{EURD} instrument
\citep{bowy97.apj.eurd}
%

Diffuse FUV spectrograph performance is a function of
optical quality, dispersion, efficiency, and scattering 
due to surface or grating imperfections.
The collecting mirrors were measured to have a focal line width 
of  \texttt{<} 100 \ensuremath{\mu}m FWHM at the focal distance of 15 cm
and a surface roughness of 10 {\AA} RMS.
The gratings
were polished to \texttt{<} 35 {\AA} RMS surface roughness
and holographically ruled with a blazed  profile
formed by ion-beam ablation of chemically etched grooves,
providing 65\% peak groove diffraction efficiency.  
The Long and Short channels (hereafter referred to as ``L/S'') 
use the same grating figure
while the Long channel works in first order at 3000 line mm$^{-1}$
and the Short channel works in second order at 2250 line mm$^{-1}$.
The L/S optics use coatings, MgF$_{2}$ on Al and 
B$_{4}$C  on a thin Cr base \citep{keski95.spie.bc},
were chosen to optimize bandpass efficiency and 
resist degradation by atomic oxygen. 

Care was taken to reduce background noise from strong airglow lines, stars, detector 
and ion background, and optics fluorescence. 
The Long channel includes a flexure-mounted, zero-power CaF$_{2}$ cylindrical meniscus filter 
that excludes geocoronal Lyman \ensuremath{\alpha} before the 150 ${\micron}$ slit.
The Long  band response is limited at short $\lambda$ by the CaF$_{2}$ filter
and at long $\lambda$ by the detector photocathode (see below)
such that high-order diffraction is rejected.
The Short band is un-filtered and its response is limited by 
falling optical coating and diffraction efficiency at short $\lambda$,
and by the detector photocathode at long $\lambda$, although
high-order response to the bright He {\sc I} $\lambda$ 537 is found in flight data.
The Short channel uses MgF$_{2}$ windows that transmit
$\lambda >$1150 \AA to detector areas adjacent to the science field 
so that instrumentally scattered  Lyman $\alpha$ airglow
can be monitored. 
Consequently, the L/S science fields of view are 
7.4$^{o}$ x 4.3' and 4.0$^{o}$ x 4.6'.
We estimate the radiation induced filter fluorescence from
the filters' fluorescence rate due to a predicted orbital radiation environment, the 
detector's sensitivity to the fluorescence spectrum, and 
the detector's view of the filters. 
The shutter can be selected to admit 0\%, 1\%, 10\% 
or 100 \% of the available light for safe and photometric observations 
of bright sources.

\subsection{Sensor and Systems}

Each channel's spectrum is focused upon a separate position-sensitive 
photon-counting microchannel plate (MCP) Z-stack that is top coated 
with an L/S optimized photocathode of CsI and KBr.
The stacks' 23mm square active fields share
a single event position encoding system 
by using a unique cross delay-line anode
\citep{jgr02.jass.tdc}
with a bifurcated line to sense the spectral axis
and a continuous line to sense the imaging axis. 
The anode axes are rotated 
15$^{\circ}$ with respect to the dispersion plane to mitigate the 
appearance of false spectral features from
anode or electronics differential non-linearity. 
Event axes positions are independently determined 
by precisely timing the arrival times of amplified anode pulses.
The position conversion system has a fixed dead-time of 86 $\mu$s per event.
A stimulation unit injects 
an artificial event signal for each field corner at 10 Hz so that
thermal drifts in the position encoding system can be calibrated 
in flight. The amplifiers also produce a signal proportional 
to every event's charge amplitude to allow for the rejection of low 
amplitude noise or high amplitude ion and cosmic ray events. 
A count-rate monitor turns off the detector in case of excessive count-rates.
Observation restart is automatically attempted within 2 to 30 s
so that entire survey sweeps will not be lost to momentary bright stars transits.
Timing marks are inserted into the data stream to 
accurately account for such interruptions.

The spectrographs are contained in an enclosure including the gratings, 
order baffling, the detector, a shutter-slit unit, 
a mirror unit with field baffling, and a deployable 
dust cover. Each channel is optically baffled from the other. 
Thermal ion detector noise is suppressed
by positively biased wire grids and baffles on all enclosure openings
and the detector face, and by high-energy magnets at the slit. 
The  5x8 cm optics are bonded to thermally-matched metal flexures
using a low-modulus adhesive, and then attached to three-point mounts.
The gratings were aligned with divergent FUV radiation illuminating the slit 
and mirror-slit alignment used visible collimated light. 
Full-system ground calibration with collimated FUV light established
the field width and angular scale.
The complete 22 kg instrument is 45$\times$45$\times$15.5 cm in size. 

The instrument 
uses ground command
to set payload operating modes and to tune critical engineering 
parameters such as detector voltage levels, detector electronics 
thresholds, and shutter timing. 
On-board software controls the instrument operation, data packetization and flow,
and  spacecraft-time synchronization. 
Commands can synchronize the instrument and spacecraft clocks to 250 ms.
Detector events (photons, stimulation events, noise events) are 
queued into packets interleaved with a 10 Hz time reference  and other marks
used during data reduction for attitude synchronization.
Science data are transmitted to a mass memory system at 200 kbps for downlink. 
The data system throughput impacts the event rate
resulting in a net dead-time of $\sim$133 $\mu$s per event, 
corresponding to a throughput loss of 5\% for a 1kHz event rate.
Dead-time was determined by
on-orbit observations of identical bright sky regions using the 10\% and 100\% 
shutter positions.

\section{Photon Event Processing and Calibration}

Individual photon data events are subject to automatic processing 
in-orbit and to ground-based pipeline processing. 
Photon events are selected for a valid 
pulse height based upon the nominal distribution (65\% FWHM). 
Low-amplitude noise events are discarded by the on-board electronics.
Typically, $\sim$7\% of telemetered events,
presumably caused by ions and cosmic rays,
are discarded due to excessive pulse amplitude.
Photon events are marked with the concurrent total count rate 
so that dead-time corrections can be calculated.
Photon detector coordinates are pipeline
corrected for detector electronics thermal drift and distortion 
every 5 s using a 2-d gradient distortion referenced 
to the stimulation marks' centroids. The drift correction, almost entirely in
the dispersion direction, decays from 
\ensuremath{\sim}0.6 {\AA} to insignificance within 120 s of an 
observation's start. The stim centroid positions are stable to \ensuremath{\sim}0.15 {\AA}. 
Photon events occurring outside of an active boundary, defined 
using deep detector exposures, are rejected including a 1 -- 10 
s$^{-1}$ event ``warm-spot'' near the Long channel's detector edge. Other 
photon data integrity checks reject \texttt{<} 0.5\% of the data.

Photon wavelength, \ensuremath{\lambda}, 
and field angle, \ensuremath{\phi},  are derived from detector coordinates 
using simple polynomial plate scales 
in the spectral and angular dimensions. 
The imaging angular scale, 0.31$^{\circ}$ $mm^{-1}$, was measured pre-flight 
using a collimated
FUV spectral illumination at precise field angles.
Imaging L/S angular resolution, averaged over the bandpass,
was measured from bright-star crossings and are 6.5' and 4.5' HEW.
The  L/S  spectral dispersion scales, 17.9 {\AA} $mm^{-1}$ and 12.0 {\AA} $mm^{-1}$,
were determined by comparing the Gaussian-fit centroids
of  measured auroral emission lines to centroids derived from 
predicted air-glow emissions \citep{auric}
that have been smoothed by a 2 {\AA} HEW Gaussian. 

A $\lambda$ distortion correction 
was constructed using composite emission line spectra from 20 orbits 
of auroral observation. The L/S central $\lambda$ error correction 
for each of 15 and 9 lines was determined at 45' intervals.
The correction applies a 2-d gradient to $\lambda$ and $\phi$
with a residual angle-integrated line $\lambda$ centroid error
of  0.5 {\AA} and 0.75 {\AA}.
The resulting spectra shows remaining higher order distortions.
Consequently, a second $\lambda$ correction 
was constructed using composite line spectra from 155 orbits 
of auroral observations. The L/S $\lambda$ error-correction 
for 22 and 15 lines was determined in 5' angular steps and then
boxcar smoothed at 15' angular width. 
Corrections for $\lambda$'s between the measured lines
were estimated by a low-tension spline fit. 
The second correction improves spectral resolution and centroid accuracy,
but introduces flat-field non-linearity as described below.
The result is an L/S angle-integrated median 
$\lambda$ centroid errors of  0.11{\AA} and 0.081 {\AA},
with mean errors of 0.24 $\pm$ 0.41 {\AA} and 0.21 $\pm$ 0.24 {\AA}.
Residual systematic errors of $\sim$1 \AA\, 
 which we intend to correct in future work,
persist in the S channel for $\lambda <$1000 \AA.

Spectral resolution is measured from the observed auroral spectra,
together with airglow spectra for the S channel,
after applying the distortion corrections. 
Because the sources are diffuse, the results are independent of spacecraft pointing knowledge.
The first correction gives an
L/S spectral half-energy line width, 
averaged over the angular field,
of 3.2  {\AA} and 1.9 {\AA}
while the second correction widths are
 2.95 $\pm$ 0.76  {\AA} and 1.71 $\pm$ 0.33 {\AA}.
These values are upper limits to the resolution because any
intrinsic width of multiplets in the source spectra were not accounted.

Detector differential non-linearity (DNL) arises from the $\lambda$ correction 
process from its application to distributed noise sources
such as intrinsic detector noise, events from ions and penetrating cosmic rays,
or instrumentally scattered airglow lines,
a significant noise source in the Short channel.
The  processing-induced DNL can be corrected and eliminated
if the distributed noise spectra is known because the DNL is fixed in detector coordinate space.
Whether {\em uncorrected} DNL can charade as a spectral feature depends on the
relative intensity of the distributed noise to the true external signal and on
whether both the DNL and the feature have similar spectral profiles. 
We use simulations of the induced DNL by processing 
a uniform random distribution of photons on the detector
to estimate that completely uncorrected DNL can appear as 
line emission features with an intensity of $\sim$1.5 \% 
of the distributed noise level.

An on-orbit 'dark' background integration of events 
due to intrinsic detector, particle and penetrating radiation noise
was accumulated in 42 ksec from 250 orbits that include
0\% shutter observations which were confirmed to exclude occasional 
mis-positioned shutter airglow leaks.
The detector dark background rate is from
0.02--.04 counts s$^{-1}$ \AA$^{-1}$ 
over the full science field angular height.
The dark background
shows the expected  $\lambda$ correction DNL features.
In the S channel, instrumentally-scattered Lyman $\alpha$ 1216 {\AA}
airglow radiation is a significant contributor to the intrinsic background.  
The S channel scattering function, measured by 
comparing integrated closed-slit and open-slit background observations,
is well fit by an exponential with 190 {\AA} width and
an intensity of 0.016 counts s$^{-1}$ \AA$^{-1}$ at the band center.
The airglow correction can be estimated
for observations subject to time-variable airglow
by using the Short channel MgF$_{2}$ filter regions to
measure the relative intensity of scattered Lyman $\alpha$ radiation.

\section{Sensitivity Calibration}

The sensitivity to diffuse radiation is determined by the grasp, \ensuremath{\Gamma}, 
a product of the solid-angle, $\Omega$, and effective area, A$_{eff}$.
The solid angle for L/S, a product of the slit's angular width 
and the viewed angle per $\lambda$,
constrained by the detector boundary,
are  $1.6\times 10^{-4}$\,sr
and  $8.4\times 10^{-5}$\,sr, respectively.
The A$_{eff}$ was determined using special calibration observations
where repetitive roll-sweeps of the 4.5' field of view crossed a star,
and using regular sky-survey operations. 
Data were acquired over a range of field-angle positions on the detector
for both cases.
For the calibration data, a 'timed-sweep' method was used to derive star exposure times
by dividing the spacecraft roll rate by the ground-calibrated angular slit width. 
About 15\% of the sweeps were rejected due to data faults.
The net stellar signal was determined after 
subtracting adjacent sky background 
({$\sim$}5--10{\%} of the stellar flux).
The sweeps' stellar count rates show a nearly
Gaussian distribution with a $\sim$15\% width.
For the sky-survey data,
we used the sky mapping method \citep{je:05apj_mission}
to obtain source and background spectra of the field stars.
The sky-map derived flux was verified to reproduce the more 
comprehensive timed-sweep calibration results to within 10--15\%.

For the L channel, calibration observations of
G191B2B and HD188665 provided 253 s and 19 s of stellar exposure,
and 13.2 k and 26.1 k photons (background subtracted) for
comparison to the spectra of \cite{kruk95} and \cite{buss95}, respectively.
The field stars HD 74753, HD 74273, and HD 72014
were mapped and referenced to their $IUE$ spectra.
The A$_{eff}$ for each star was derived by dividing
the observed count rate by the reference flux,
with a typical statistical error per 1.5\AA\, spectral bin of $\sim$10\%.
The stars' A$_{eff}$ were averaged using inverse 
counting-error weighting and fit with a third-order polynomial 
to obtain the spectral calibration curve, shown in Fig. 2.
The individual star's A$_{eff}$ deviate from the 
weighted average by $\sim $20--30\% over the band
with no discernible in-flight time trend.
At the L band center, we find
A$_{eff}$= 0.20 cm$^{2}$,
corresponding to a full-field  grasp of
$\Gamma = 3.2\times 10^{-5} \, $cm$^{2} $sr$^{-1}$.
%

For the S channel, 
a calibration observation of HD 93521,
conducted March to April 2004, 
provided 146 s and 9300 photons (background subtracted)
for comparison to the spectra of \cite{buss95}.
Also, the field star HD 1337 was
observed in the survey during November 2003 
for 57 sec with 3500 counts 
and referenced to its $HUT$ spectra.
Each stars' A$_{eff}$ was derived using eight $\lambda$ 10--30
bins that avoid the airglow Lyman series
and have $\lesssim$ 6\% statistical error, 
except for the $\lambda$ 915-925 bin with a 37\% error.
The A$_{eff}$ bins were fit with a second-order polynomial
to obtain spectral calibration curves, shown in Fig. 2.
At the S band center we find that
A$_{eff}$= 0.035 cm$^{2}$ for HD 1337, and
A$_{eff}$= 0.016 cm$^{2}$ for HD 93521.
There appears to be a significant S-channel degradation with time.
The earlier A$_{eff}$
corresponds to a full-field height S grasp of
$3.0\times 10^{-6} \, $cm$^{-2} $sr$^{-1}$.
HD 1337 (AO Cas) is an eclipsing binary with a 3.4 day period
and a $\sim \pm$7.5\% photometric variation in FUV resonance lines,
such as N{\sc V}Ê$\lambda\lambda$ 1240 and C{\sc IV}Ê$\lambda\lambda$ 1550 
\citep{AOcas}. Our measurement of AO Cas should be of
a mean intensity because it broadly samples the binary phase in 
$\lambda$ bins that are large compared to observed phased velocity shifts.
We intend to refine our sensitivity calibration and its temporal behavior
in future work by measuring more stars.

Systematic errors in the determination of exposure times and in the 
reference stellar spectra are likely to dominate our flux calibration accuracy.
We estimate these errors systematic overall to be $\sim25 \%$.
The relative angular sensitivity variation from the mean response is 6.5\%
as found from the off-axis response to bright in-band telluric 
diffuse night glow emission lines (e.g. O I 1356{\AA}, H I 1026{\AA}) 
accumulated over all observations. We presume that the
night glow lines are effectively constant in intensity over the sky because these data
are taken over a wide range of zenith angles.
Compared to its pre-launch performance,
{\em SPEAR's} on-orbit sensitivity was unfortunately degraded from
its delivered performance 
due to contamination caused by spacecraft and launch ground operations.

\section{Conclusion: Performance}

{\em SPEAR's} sensitivity can be estimated from the on-orbit performance values.
The ultimate capability to detect diffuse emission with {\em SPEAR} 
depends on mission factors such as sky exposure, attitude knowledge and
background temporal behavior.  These factors are treated elsewhere 
in this issue \citep{je:05apj_mission}.
We find that the dominant cause of background is 
a combination of detector dark noise and instrumentally scattered airglow 
for the Short channel, and dark noise and 
interstellar continuum for the Long channel.
In-orbit measured L/S background rates have typical values of  
$\sim$0.02 counts s$^{-1}$ {\AA}$^{-1}$
for diffuse high latitude sky fields (excluding stars). 
These values correspond to an L/S isolated emission line 3-$\sigma$ sensitivity
of 55  and 750 photons s$^{-1}$ cm$^{-2 }$ sr$^{-1}$
for diffuse illumination filling the full angular field,
presuming a 10 ksec observation, and a background 
determination over 100{\AA} of bandpass.

Limits to the measurement of emission lines that are 
not fully isolated from airglow lines, such as
OVI 1032 {\AA} near Lyman $\beta$ 1027 {\AA} and
CIII 977 {\AA} near Lyman $\gamma$ 972 {\AA} ,
depends on the intensity of the airglow line and the accuracy of spectral profiling
used in compound spectral line-fit modeling. 
We leave a systematic treatment
of such modeling  and its limitations to papers 
specifically observing these spectral lines. 
As an example of the sensitivity, however, we provide that 
7000 LU of O {\sc VI} $\lambda$ 1032 was measured with 5 $\sigma$ significance
in a 31 ks observations toward the Eridanus Loop \citep{kregenow:05apj_eri}.

\acknowledgments
\emph{SPEAR / FIMS} 
is a joint project of KASI \& KAIST (Korea) and U.C., Berkeley (USA),
funded by the Korea MOST and NASA Grant NAG5-5355.
We thank the team for their dedication in producing the instrument,
and Jerry thanks Liz and Dan for immeasurable contributions.


%

\begin{figure}
\epsscale{1.}
\plotone{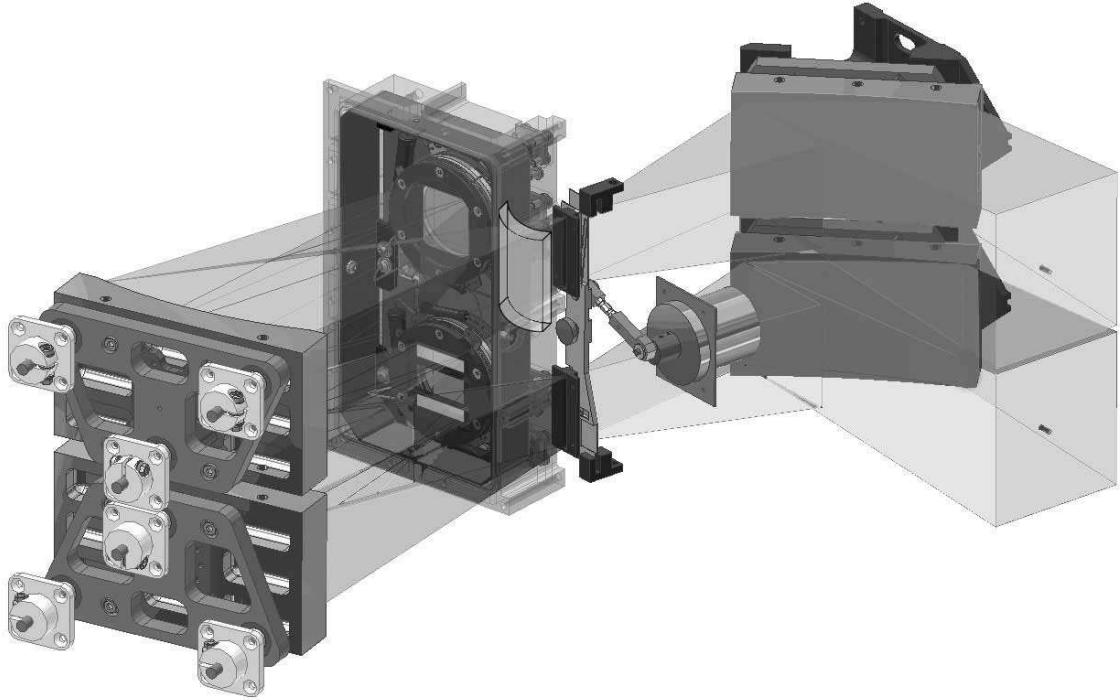}
\caption{
The {\it SPEAR} dual-spectrograph optical layout. 
Light is collected by the mirrors \emph{(right)}
to a motorized slit shutter. The Long channel \emph{(upper)}
is filtered by a CaF$_{2}$ meniscus after the slit.
The light then diffracts from gratings \emph{(left)} 
to photon counting detectors.
The Short channel \emph{(lower)} is un-filtered, 
although MgF$_{2}$ windows on the detector
abut the science field to monitor airglow.
For scale, the grating is $\sim$8 cm long.
 }
\end{figure}

\begin{figure}
\epsscale{0.8}
\plotone{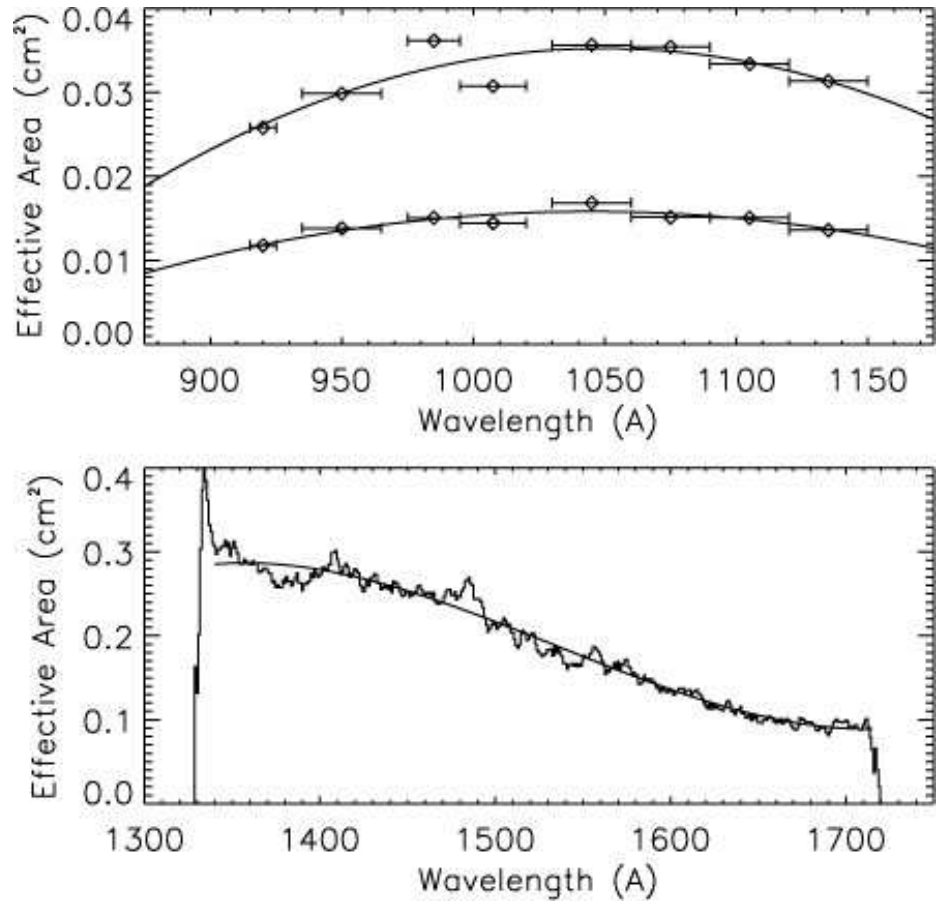}
\caption{
The {\it SPEAR} A$_{eff}$ calibration fit to stellar observations for 
{\it (top panel)} the Short band  
in November 2003 {\it (upper curve)} and April 2004 {\it (lower curve)},
and for the Long band
 {\it (bottom panel)}  fit to the average histogram from five stellar observations. 
\label{fig2} }
\end{figure}

\end{document}